\documentclass[final,3p,times]{elsarticle}

\usepackage{amssymb}

\usepackage{amsmath}
\usepackage{xcolor}
\usepackage[draft]{changes}

\journal{Physica A}

\begin{document}

\begin{frontmatter}



\title{Polarization-induced stress in the noisy voter model}


\author{Miguel Aguilar-Janita}
\author{Andres Blanco-Alonso}

\author{Nagi Khalil}
\address{Complex System Group \& GISC, Universidad Rey Juan Carlos, 28933 M\'ostoles, Madrid, Spain}

\begin{abstract}
  A new model for the dynamics of opinion formation is proposed and analysed at the mean-field level. It can be regarded as a generalization of the noisy voter model in which agents update their binary states by copying others and by an intrinsic mechanism affected by the degree of polarization in the system. It also takes into account whether the agents enhance or reduce their intrinsic mechanism upon increasing polarization. Four phases or shapes of the steady-state probability of a fraction of agents in a given state are found (unimodal, bimodal, W and M). In the unimodal (resp. bimodal) phase, the copying (resp. intrinsic) mechanism is globally dominant, while in the W (resp. M) phase the copying (resp. intrinsic) mechanism is the relevant one close to the consensus states while it reduces its influence as approaching coexistence. In the thermodynamic limit, the bimodal and W phases disappear, while the unimodal and M phases prevail. The theoretical results, obtained analytically from the master equation, and the numerical simulations are in good agreement. 
\end{abstract}




\end{frontmatter}


\section{Introduction}
\label{intro}

The dynamics of opinion formation emerge from the need of people to make decisions, updating their beliefs and thoughts through their own reflections or as a consequence of the observation of others. The latter is often referred to as social interaction and has been studied using different perspectives \cite{turner1988theory,moussaid2013social,toscani2006kinetic,khalil2021approach}. Of particular interest is the approach based on Statistical Physics of classical systems, which has provided many instances of surprising emergent phenomena common to other complex systems \cite{castellano2009statistical,jkedrzejewski2019statistical}.

In the simplest situation, people hold two possible opinions on a given topic, changing them by blindly copying others. This scenario is modeled by the Voter Model (VM) \cite{clifford1973model,holley1975ergodic}, which proposes a simple stochastic dynamics: a randomly chosen agent copies the state of a neighbor, also selected at random. Despite its simplicity, the dynamics of opinion formation turns out to be nontrivial, depending significantly on the system size and the topology of the social interaction network \cite{suchecki2005voter,vazquez2008analytical}. 

Taking a step beyond the VM, the noisy voter model (NVM) \cite{kirman1993ants,carro2016noisy,peralta2018stochastic,perachia2022noisy,pymar2021stationary} introduces an intrinsic or noisy mechanism for opinion change. Once an agent is randomly selected, her opinion changes with some probability \(p\), while a VM update is implemented with probability \(1-p\). In a well-mixed population, the NVM describes a finite-size transition from a bimodal to a unimodal phase as \(p\) increases. For \(p\) below a critical value \(p_c\), the copying (herding) mechanism dominates over the noise, maintaining the system close to consensus states (bimodal phase). Conversely, for \(p>p_c\), the balance shifts, and the most probable configuration becomes one of coexistence, where the same fraction of people holds opposite opinions (unimodal phase). As the number of agents tends to infinity (thermodynamic limit), the critical probability \(p_c\) tends to zero, and the unimodal-bimodal transition disappears.

The NVM incorporates the main mechanisms needed to construct more specific models. Thanks to this, the NVM has been adapted to include external information in the context of financial markets \cite{carro2015markets}, the effect of zealots and contrarians \cite{mobilia2007role,khalil2018zealots,khalil2019noisy}, aging and latency \cite{artime2018aging,artime2019herding,peralta2020reduction,peralta2020ordering,palermo2023spontaneous,llabres2023aging}, more than two opinions \cite{herrerias2019consensus,khalil2021zealots,ramirez2022local}, and fluctuating environments \cite{kudtarkar2023noise,caligiuri2023noisy}, just to mention a few examples. Interestingly, in these and other examples new phases typically emerge. This raises the following question: is the picture depicted by the NVM robust against slight modifications of the herding and/or noise mechanisms?

The question has been partially addressed with respect to the copying mechanism by considering non-linear interactions in the so-called nonlinear noisy voter model \cite{peralta2018analytical}. There, the probability of an agent to copy an opinion is taken as the \(\alpha\)--power of the fraction of agents holding that opinion, with the case \(\alpha=1\)  corresponding to the NVM. It turns out that the outcome of the model in a well-mixed population depends critically on 
\(\alpha\): for \(\alpha<1\), only the unimodal phase is present; for \(\alpha\in(1,5)\), both unimodal and bimodal phases can be reached depending on the noise intensity; and for \(\alpha>5\), an additional (trimodal) phase emerges. The trimodal phase is characterized by a symmetric probability function having two global minima at the consensus states, a local minimum at coexistence, and two global maxima at intermediate values. Importantly, in contrast to the case  \(\alpha=1\) (NVM), all transitions for \(\alpha>1\) persist in the thermodynamic limit.

In this work, we study the effect of a modification of the noise term on the NVM, without perturbing the herding mechanism, in a well-mixed population. In doing so, we aim to capture the social pressure (stress) that may be present when an agents invoke the intrinsic mechanism of opinion update. Particularly, the noise term is corrected by a term proportional to the fraction of agent in opposite states (fraction of active links), this way preserving the symmetry between the two opinions. We assess the influence of the new term on the different phases of the system as well as the  finite or thermodynamic nature of the transitions. 

The reminder of the paper is organized as follows. Next section is devoted to describe the model, where we also observe that the proposed model can be seen as the leading order of a more general one, in which the noise and herding terms can be (almost) any functions of the state of the system. The main theoretical results are included in Secs.~\ref{steadystates} and \ref{phasediagram} where we first obtain relevant properties of the steady-state probability function for number of agents holding a given opinion and then construct a phase diagram with the different phases as we change the relevant parameters of the system. Most results are exactly obtained from the master equation and are compared against molecular simulations in Sec.~\ref{numericalsimulations}. Finally, the work ends with a discussion.

\section{Model\label{model}}

\subsection{State}
The system has \(N\) agents, each one holding one possible opinion or state, \(0\) or \(1\). From a macroscopic point of view the state of the system is given by the number of agents with a given state. In particular, we use \(n\) to denote the number of agents with state \(1\). Alternatively, we also consider the magnetization,
\begin{equation}
  \label{eq:magnetization}
  m=2\frac{n}{N}-1,
\end{equation}
which is an intensive magnitude taking values between \(-1\) and \(1\).

Two consensus states can be identified: all agents holding state \(0\), for which \(n=0\) and \(m=-1\), and all agents with opinion \(1\), now \(n=N\) and \(m=1\). When \(N\) is an even number, pure coexistence corresponds to \(n=\frac{N}{2}\) and \(m=0\).

\subsection{Dynamics}

The dynamics is driven by two social mechanisms: herding and noisy-like terms. The former takes into account the tendency of agents to copy others' opinions, as in the Voter Model (VM): the probability of an agent to change her opinion is taken as proportional to the number of agents with a different state. The noise term takes into account the intrinsic tendency of changing opinion: even if the system is in a consensus state, an agent can still change its opinion with some probability. In our model, this probability is affected by the degree of polarization, which is assumed to be proportional to \(n(N-n)\).

More precisely, we assume that the probability function \(P(n,t)\) of finding the system in state \(n\) at time \(t\) obeys the continuous-time master equation
\begin{equation}
  \label{eq:master}
  \partial_t P(n,t)=(E^+-1)\pi^{-}(n)P(n,t)+(E^--1)\pi^{+}(n)P(n,t)\;,
\end{equation}
where the operator \(E^\pm\) increases (+)/decreases (-) the argument of any function of \(n\) by one and the rates \(\pi^{\pm}(n)\) are associated to the transitions \(n\to n\pm 1\). The latter are given by 
\begin{eqnarray}
  \label{eq:pimasn}
  && \pi^+(n)=(N-n)\left[a+b\frac{n(N-n)}{N^2/4}+h\frac{n}{N}\right]\;,\\
  \label{eq:pimenosn}
  && \pi^-(n)=n\left[a+b\frac{n(N-n)}{N^2/4}+h\frac{N-n}{N}\right]\;,
\end{eqnarray}
where \(a\ge 0\) accounts for the intrinsic change of opinion, \(b\) is a positive or negative coefficient that modifies the polarization contribution, and \(h\ge 0\) tunes the copying or herding mechanism.

The sign of \(b\) provides two different social behaviours. While for \(b>0\) agents react positively to the polarization, by enlarging their intrinsic ability to change opinion, for \(b<0\) we have the opposite effect. The case of \(b>0\) is reminiscent of what actually happens for a system affected by the same number of opposite zealots \cite{khalil2018zealots}. However, we stress that in the present model, as opposed to the case of zealots, the stress mechanism is affected by the state of the system, being zero for the consensus states and maximum for coexistence.  

Since the rates are non-negative functions, \(\pi^\pm(n)\ge 0\) for \(n=0,\dots, N\), not all values of \(b\) are possible. To find the possible values of  \(b\), it is enough to impose \(\pi^+(n)\ge 0\), thanks to the symmetry \(\pi^+(N-n)=\pi^-(n)\). After a direct calculation, we get
\begin{equation}
  \label{eq:ratepos}
  \frac{b}{h}\ge -\frac{1}{4}\left[1+\frac{2a}{h}\left(1+\sqrt{1+\frac{h}{a}}\right)\right]\;,
\end{equation}
which is independent of \(N\). 

Finally, it is worth noting that the model reduces to the NVM for \(b=0\). If, in addition, we set \(a=0\), then we recover the VM. In this work, however, we always consider that the inequality (\(>\)) of Eq.~\eqref{eq:ratepos} is fulfilled, even if \(h=0\) (no herding), in order to avoid possible absorbing states (where the system could get trapped).

\subsection{Generalization}

Our model can also be interpreted as an approximation of more general models. To see this, consider the rates of Eqs.~\eqref{eq:pimasn} and \eqref{eq:pimenosn} written in term of the magnetization as 
\begin{eqnarray}
  \label{eq:pi+m}
  && \pi^+(m)=N\frac{1-m}{2}\left(c_0+c_1m-c_2m^2\right)\;, \\
  \label{eq:pi-m}
  && \pi^-(m)=N\frac{1+m}{2}\left(c_0-c_1m-c_2m^2\right)\;,
\end{eqnarray}
where
\begin{eqnarray}
  \label{eq:c0}
  c_0&=&a+b+\frac{h}{2}\;, \\
  c_1&=&\frac{h}{2}\;, \\
  \label{eq:c2}
  c_2&=&b\;.
\end{eqnarray}
In this way, the rates has the form of the first orders of a Taylor expansion of a more general ones:
\begin{eqnarray}
  \pi^+(m)&=&N\frac{1-m}{2}f(m)\;, \\
  \pi^-(m)&=&\pi^+(-m)\;,
\end{eqnarray}
with
\begin{eqnarray}
  f(m)=c_0+c_1m-c_2m^2+\dots
\end{eqnarray}
Here, the coefficients $c_i$ can take any possible values that fulfill the condition \(f\ge 0\) for all \(m\), not necessarily those given by Eqs.~\eqref{eq:c0}--\eqref{eq:c2}. 
The study of this generalization is very interesting but beyond the scope of the present work. 

\section{Steady states\label{steadystates}}

\subsection{Steady-state probability function}

The different behaviors of the system are encoded in the form of the steady-state probability \(P(n)\). The latter can be obtained by setting \(\partial_tP=0\) in Eq.~\eqref{eq:master}:
\begin{equation}
  \label{eq:stmaster}
  (E^+-1)\pi^{-}(n)P(n)+(E^--1)\pi^{+}(n)P(n)=0\;,
\end{equation}
for \(n=0,\dots,N\), with the boundary conditions
\begin{equation}
  P(-1)=P(N+1)=0\;,  
\end{equation}
and the normalization restriction
\begin{equation}
\label{eq:suma}
  \sum_{n=0}^NP(n)=1\;.
\end{equation}

The problem \eqref{eq:stmaster}--\eqref{eq:suma}, with the rates in Eqs.~\eqref{eq:pimasn}--\eqref{eq:pimenosn} being positive, has a unique solution, as can be seen as follows. For \(n=0\), using the boundary conditions and the fact that \(\pi^-(0)=\pi^+(N)=0\), we get
\begin{equation}
  \label{eq:sol1}
  P(1)=\frac{\pi^+(0)}{\pi^-(1)}P(0)\;.
\end{equation}
Now, taking \(n=1\) and using the previous relation, we arrive at
\begin{equation}
  P(2)=\frac{\pi^+(1)}{\pi^-(2)}P(1)\;.
\end{equation}
Proceeding the same way, we have in general
\begin{equation}
  \label{eq:solg}
  P(n)=\frac{\pi^+(n-1)}{\pi^-(n)}P(n-1)\;,
\end{equation}
for \(n=0,\dots, N\).

\subsection{Properties of \(P(n)\)}

An important consequence of Eq.~\eqref{eq:solg} is that, for the parameters of the system verifying the inequality (\(>\)) at Eq.~\eqref{eq:ratepos}, the steady-state probability is
\begin{equation}
  P(n)>0\;,
\end{equation}
for all \(n\). This relation holds because, on the one hand, under the previous assumptions we have  \(\pi^\pm>0\) for all \(n\) and, on the other hand, the probability function is normalized.  

Another important property of \(P(n)\) is
\begin{equation}
  \label{eq:symm}
  P(N-n)=P(n)\;.
\end{equation}
This can be seen by writing Eq.~\eqref{eq:stmaster} as 
\begin{equation}
  (E^--1)\pi^{-}(N-n)P(N-n)+(E^+-1)\pi^{+}(N-n)P(N-n)=0\;,
\end{equation}
and observing that
\begin{equation}
  \pi^\pm(N-n)=\pi^\mp(n)\;,
\end{equation}
which gives rise to the following problem 
\begin{equation}
  (E^--1)\pi^{+}(n)\tilde P(n)+(E^+-1)\pi^{-}(n)\tilde P(n)=0\;,
\end{equation}
with \(\sum_{n=0}^N \tilde P(n)=1\) and \(\tilde P(n)=P(N-n)\). Since this new problem coincides with the one given by Eq.~\eqref{eq:stmaster}, and thanks to the uniqueness of the solution, it is \(\tilde P(n)=P(N-n)=P(n)\).

\section{Phase diagram\label{phasediagram}}

\subsection{Critical lines}

A phase transition takes place when \(P(n)\) loses or gains a maximum or a minimum. We begin with the simplest cases of the central point\footnote{From now on, in order to simplify the study, we assume an even number of agents \(N\).} \(n=N/2\) and the extreme ones \(n=0,\, N\). Later we will see that, in fact, they are the relevant points to draw a phase diagram showing all different phases.

Taking advantage of the symmetry property of the steady-state probability \(P(n)=P(N-n)\), we directly obtain that the central point \(n=N/2\) is a maximum or a minimum. Hence, a phase transition due to the central point occurs when 
\begin{equation}
  \label{eq:condn0}
  P(N/2-1)=P(N/2)=P(N/2+1)\;.
\end{equation}
Using this condition with Eq.~\eqref{eq:stmaster} we obtain
\begin{equation}
  \label{eq:tr1}
  b=\frac{N^2}{N^2-2N-4}\left(\frac{h}{N}-a\right)\simeq \frac{h}{N}-a\;,
\end{equation}
where the approximate expression holds for \(N\gg 1\). Note that for \(b=0\) we recover the critical point of the NVM.

As for the extreme values, for symmetry reasons it is enough to consider \(n=0\). A change from a minimum to a maximum and vice versa at \(n=0\) takes place when \(P(0)=P(1)\). Using this condition with Eq.~\eqref{eq:sol1} and the form of the rates, we arrive at 
\begin{equation}
  \label{eq:tr2}
  b=\frac{N^2}{4}\left(a-\frac{h}{N}\right)\;.
\end{equation}
Again, for \(b=0\) we recover the NVM transition. In this case \(P(n)\) is completely flat at the transition point.

\subsection{Maxima and minima} \label{sect:max_min}

In order to find other possible transitions, and even to asses the continuous or discontinuous nature of them, we have to find the location of the maxima and minima of \(P(n)\). A maximum of \(P(n)\) is located at \(n=n_0\in\{1,\dots,N-1\}\) provided that the following conditions hold
\begin{equation}
  \label{eq:maximo}
  P(n_0-1)< P(n_0)> P(n_0+1)\;.
\end{equation}
The condition for a minimum is obtained by changing the two inequality signs.

Using Eq.~\eqref{eq:solg} with Eq.~\eqref{eq:maximo} we find the following conditions for a maximum \(m_0=2n_0/N-1\)
\begin{eqnarray}
  \label{eq:max1}
  &&\left(\frac{1}{N}+m_0\right)\left[a+\frac{(1-m_0^2)N^2-2(1+m_0)N-4}{N^2}b-\frac{h}{N}\right]>0 \;,\\
  \label{eq:max2}
  &&\left(\frac{1}{N}-m_0\right)\left[a+\frac{(1-m_0^2)N^2-2(1-m_0)N-4}{N^2}b-\frac{h}{N}\right]>0\;.
\end{eqnarray}
It is obvious that if a given \(m_0\) verifies the set of equations, then \(-m_0\) also does. Then, taking into account Eqs.~\eqref{eq:max1} and \eqref{eq:max2} and the fact that \(|m_0|\ne 1/N\) (see definition in Eq.~\eqref{eq:magnetization}), there are two possibilities:
\begin{itemize}
\item \(m_0=0\):
  \begin{equation}
    b>\frac{N^2}{N^2-2N-4}\left(\frac{h}{N}-a\right)\;.
  \end{equation}
  When the inequality is not verified, we have a minimum at \(m_0=0\) or a transition point [as given by Eq.~\eqref{eq:tr1}].
\item \(1>m_0>\frac{1}{N}\) or \(-1<m_0<-\frac{1}{N}\):
  \begin{eqnarray}
    \label{eq:l>m0}
    &&l_>(m_0)=a+\frac{(1-m_0^2)N^2-2(1+|m_0|)N-4}{N^2}b-\frac{h}{N}>0\; ,\\
    \label{eq:l<m0}
    &&l_<(m_0)=a+\frac{(1-m_0^2)N^2-2(1-|m_0|)N-4}{N^2}b-\frac{h}{N}<0\;,
  \end{eqnarray}
  where $l_>$ and $l_<$ are defined by the first equalities. For \(m_0\) to be a minimum we have to change each of the inequality signs, as usual. When the two expressions are equal to zero, then \(a=\frac{h}{N}\) and \(b=0\) for any \(m_0\).
\end{itemize}

Let us analyze the locus of the maxima of \(P(n)\) for \(m_0>1/N\) and for an even number \(N\) of agents more carefully. Under these conditions, the smallest \(m_0\) is \(\frac{2}{N}\). A maximum of \(P(n)\) is at \(\frac{2}{N}\) for the region of the space of parameters \((a/h,b/h)\) above the line \(l_>\left(\frac{2}{N}\right)=0\) and below the line \(l_<\left(\frac{2}{N}\right)=0\), has already seen. It follows that \(l_<\left(\frac{2}{N}\right)=0\) coincides with the lines \(l_<\left(0\right)= l_>\left(0\right)=0\), which give the critical condition Eq.~\eqref{eq:tr1}, that is when a maximum at \(m_0=0\) becomes a minimum. Similarly, it is readily seen that \(l_>(m_0)=l_<\left(m_0+\frac{2}{N}\right)\), meaning that the parametric region with a maximum at \(m_0\) is attached to the one with a maximum at \(m_0+\frac{2}{N}\), the next possible value. For \(m_0=1-\frac{2}{N}\), the line \(l_>\left(1-\frac{2}{N}\right)=0\) coincides with the condition Eq.~\eqref{eq:tr2} for which a maximum reaches \(m_0=1\).

The previous analysis also provides the locus of the maxima for \(m_0<-\frac{1}{N}\), by means of the symmetry of \(P(n)\). Moreover, using a similar reasoning we can also provide conditions for the minima of \(P(n)\). In any case, what is relevant is that, upon changing the parameters \(a,\,b,\) and \(h\) smoothly, the maxima and minima of the system move from one site \(n\) or \(m\) to an adjacent one while the central maximum (minimum) can turn a minimum (maximum) giving rise to two adjacent maxima (minima). That is, no ``discontinuous'' phase transitions are present, see right plot of Fig.~\ref{fig:1} for an illustration. Finally, although the analysis has been carried out assuming an even number of agents, a similar one can be done with an odd \(N\) with minimal modifications. In paticular, the same important conclusions are obtained.

\subsection{Phase diagram}

\begin{figure}[!h]
  \centering
  \includegraphics[width=.45\linewidth]{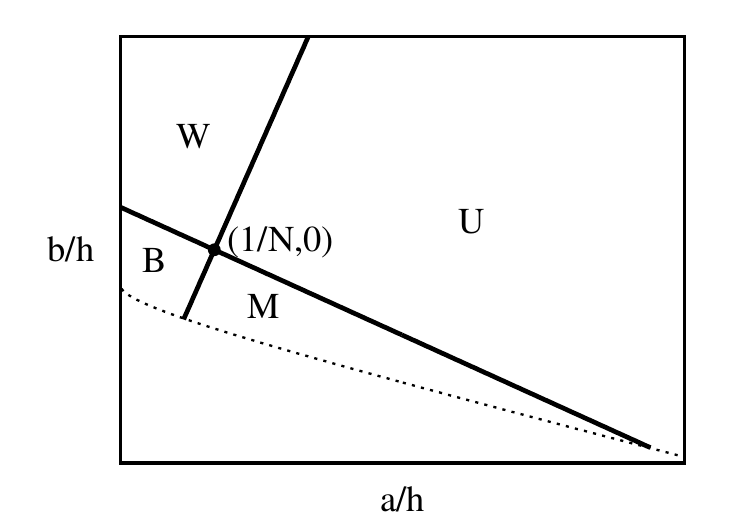}
  \hfill
  \includegraphics[width=.45\linewidth]{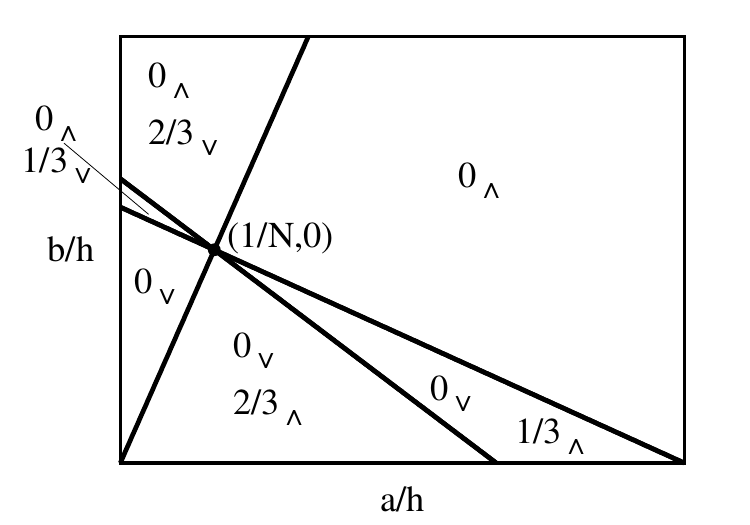}
  \caption{\textbf{Left:} Phase diagram for a system with $N$ agents. Solid lines represent the critical lines obtained from Eqs.~\eqref{eq:tr1} and~\eqref{eq:tr2} and the dashed line indicates the frontier with the forbidden region of negative transition rates, given by Eq.~\eqref{eq:ratepos}. \textbf{Right}: Example of the phase diagram for a system with $3$ agents, along with the locus of each possible value of $m_0$, showing that no discontinuous phase transitions are present in the system. The notation \(x_\vee\) and \(x_\wedge\) mean a minimum of \(P(m)\) at \(m=x\) and a maximum of \(P(m)\) at \(m=x\), respectively. }
  \label{fig:1}
\end{figure}

Using the previous results, specifically the two critical lines given Eqs.~\eqref{eq:tr1} and \eqref{eq:tr2}, we can construct a phase diagram for the form of \(P(n)\), see Fig.~\ref{fig:1}. Four different phases can be identified as \(a/h\) and \(b/h\) take different values. In the bimodal (B) phase, the system has two extreme maxima an a central minimum. In the unimodal (U) phase the probability function \(P(n)\) has one maximum at \(n=N/2\) and two minima at \(n=0,\,N\). Both phases, B and U, are present in the NVM, and can be easily identified as they include the region \(b=0\): B for \(a<h/N\) and U for \(a>h/N\). At the critical point \((a=h/N,\, b=0)\) the probability function \(P(n)\) becomes flat.

\begin{figure}
    \centering
    \includegraphics[width=\linewidth]{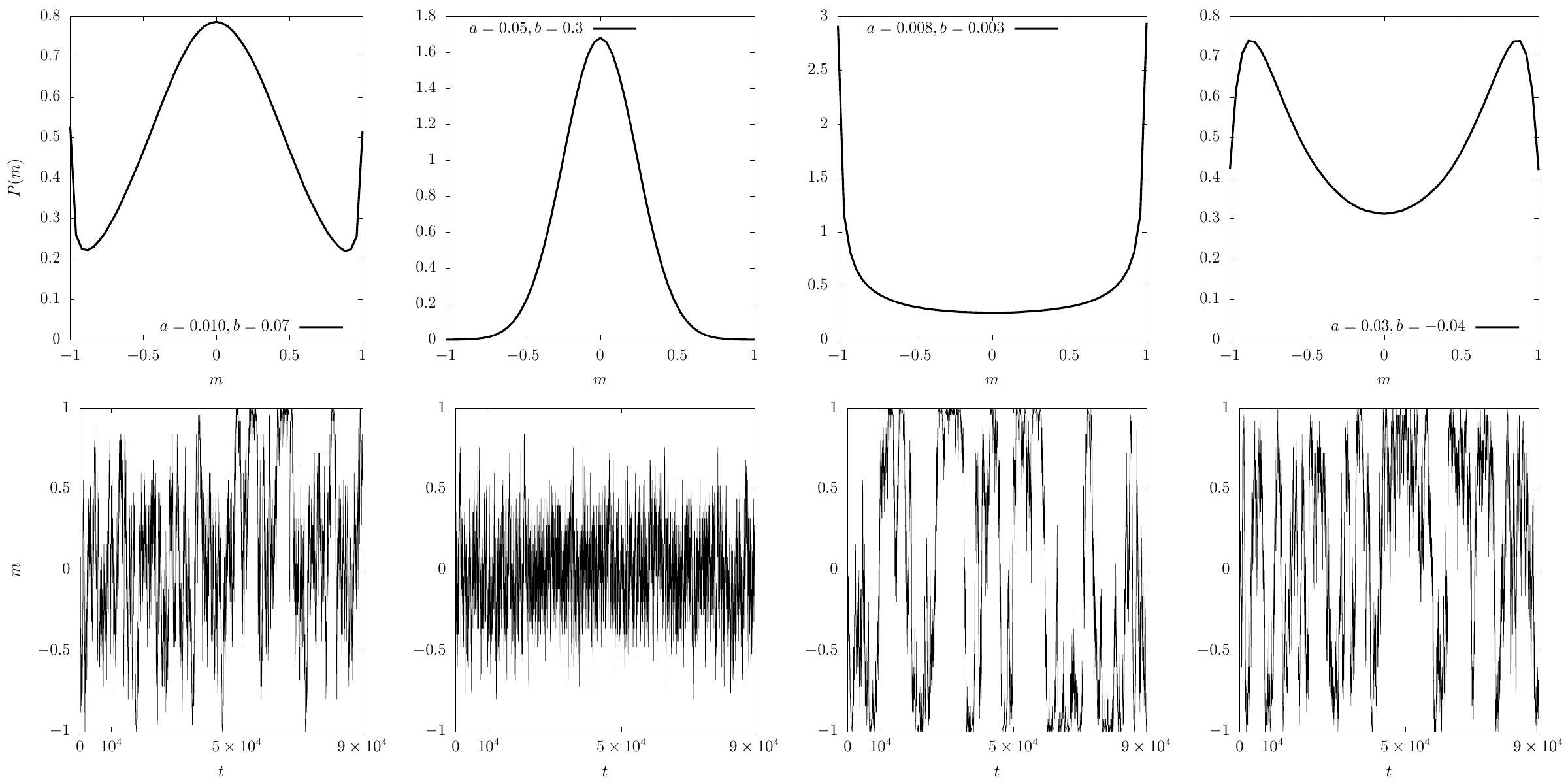}
    \caption{\textbf{Top}: Plots of the steady-state probability function \(P(m)\)  as a function of the magnetization \(m\) for all the four phases of a system with \(N=50\), obtained from numerical simulations. From left to right they correspond to W, U, B and M phases respectively. The values of $a$ and $b$ are shown in the key of each plot and $h=1$ for all of them. \textbf{Bottom}: Evolution of the magnetization of the system as a function of Gillespie's time $t$ for four systems with the same parameter as those showed in the top line.  }
    \label{fig:phases_ex}
\end{figure}

In addition to B and U, two novel phases show up: the  W phase when \(b>0\) and the M phase when \(b<0\). For fixed \(a\) and \(h\), and for \(b\) big enough, the effective noise is larger for intermediate values of \(n\), which enhances coexistence. This effect is neglected near the limiting values \(n=0,\, N\), see the form of the rates in Eqs.~\eqref{eq:pimasn} and \eqref{eq:pimenosn}. All together makes the distribution to develop three local maxima, one for \(n=N/2\) and two extreme ones for \(n=0\) and \(n=N\), all separated by two intermediate minima. Analogously, for the M phase the steady-state probability function has three local minima at \(n=N/2\), \(n=0\), and \(n=N\) separated by two symmetric maxima. An example of the \(P(m)\) for each of these phases can be seen in Fig. \ref{fig:phases_ex}. 

Note that the four phases are essentially different. On the one hand, the U and W phases are represented by an unbounded domains in the \((a/h,b/h)\) space: while the U phase can be reached for \(a/h>1/N\) or any value of \(b\), the W phase can have any \(a/h>0\) or \(b/h>0\). On the other hand, the B and M phases are bounded: for the B phase \(0\le a/h \le 1/N\)  and
\begin{equation}
  -\frac{1}{4}+\mathcal{O}\left(N^{-\frac{1}{2}}\right)=-\frac{1}{4}\frac{1}{1-\frac{2}{\sqrt{N}}+\frac{2}{N}}\le \frac{b}{h}\le \frac{1}{N}\frac{1}{1-\frac{2}{N}-\frac{4}{N^2}}=\frac{1}{N}+\mathcal{O}(N^{-2})\;,
\end{equation}
and for the M phase:
\begin{eqnarray}
  \label{eq:mb1}
  && \frac{1}{N}+\mathcal{O}\left(N^{-2}\right)=\frac{\left(\sqrt{N}-1\right)^2}{N\left(N-2\sqrt{N}+2\right)} \le \frac{a}{h}\le \frac{N^2}{8(N+2)}\left\{1+\sqrt{\frac{N+2}{N}}\left[1-\frac{2}{N}-\frac{4}{N^2}\right]+\frac{8}{N^3}\right\}=\frac{N}{4}+\mathcal{O}(1),\\
  \label{eq:mb2}
  && -\frac{N}{4}+\mathcal{O}(1)=-\frac{N}{8}\left(1+\sqrt{\frac{N}{N+2}}\right)\le b\le 0\;.
\end{eqnarray}

\subsection{Large population and thermodynamic limit}

As we increase the size of the population \(N\), for fixed values of \(a\), \(b\), and \(h\), the conditions that determine the maximum or a minimum of \(P\) at a given value of \(n\) or \(m\) become more restrictive, in the sense that a locus of a maximum or a minimum is almost determined by the parameters of the system. This is very apparent from Eqs.~\eqref{eq:l>m0} and \eqref{eq:l<m0}, for example, from which we see that \(l_>(m_0)\to l_<(m_0)\) as \(N\to \infty\). This allows us to better determine the maxima and minima of \(P\) for \(N\gg 1\) as
\begin{equation}
  \label{eq:orderp}
  |m_0|\simeq \sqrt{\frac{a+b-\frac{h}{N}}{b}}\;,
\end{equation}
which is valid for \(m_0\) not very close to \(\pm 1\). The previous expression applies for the maxima of the M phase, when \(a+b-h/N\ge 0\) and \(b\ge 0\), as well as for the minima of the W phase, when \(a+b-h/N\le 0\) and \(b<0\).

\begin{figure}[!h]
  \centering
  \includegraphics[width=.4\linewidth]{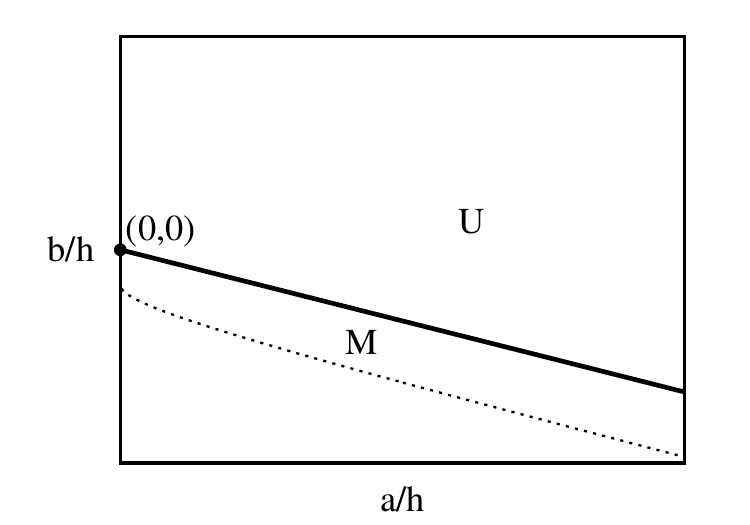}
  \caption{Phase diagram for a system with an infinite number of agents, \(N\to \infty\). In this limits, the phases W and B disappear, since the critical line of Eq.~\eqref{eq:tr2} approaches the vertical axis $a=0$.}
  \label{fig:2}
\end{figure}

Taking Eqs.~\eqref{eq:tr1} and \eqref{eq:tr2} represented in Fig.~\ref{fig:1} as a reference, it can be seen that, as \(N\) increases, the B and W phases shrink and the critical point \((a/h,b/h)=(1/N,0)\) tends to the origin. The critical line \eqref{eq:tr2} approaches the vertical axes \(a=0\) while the other critical line \eqref{eq:tr1} approaches
\begin{equation}\label{eq:cl_inf}
  a+b=0\;.
\end{equation}
For large \(a\) the line of Eq.~\eqref{eq:ratepos} becomes parallel to the previous equation, meaning that the M phase becomes unbounded. This can also be seen by taking \(N\to \infty\) in Eqs.~\eqref{eq:mb1} and \eqref{eq:mb2}. The resulting phase diagram as \(N\to \infty\) has the form shown in Fig.~\ref{fig:2}

Only the U and M phases survive in the thermodynamic limit, as shown in Fig.~\ref{fig:2}. This means that the UW, UB, MW, and BM are all finite-size phase transitions (disappearing in the thermodynamic limit), while only the transitions between the U and M phases have a thermodynamic character. Moreover, if we take \(|m_0|\) given by Eq.~\eqref{eq:orderp} as the order parameter, we see that all transitions are continuous as we change the parameters of the system, except that between B and U phases through the critical point \((a,b)=(1/N,0)\) which is discontinuous. 

\section{Numerical simulations\label{numericalsimulations}}
As a check for our analytical results, we have performed numerical simulations of the stochastic system described by the transition rates \eqref{eq:pi+m} and \eqref{eq:pi-m} using the Gillespie algorithm \cite{gillespie1976general}. All simulations have been performed for $10^8$ Gillespie steps. The main observable of our simulation is the steady-state probability function \(P(m)\) which has been computed by measuring the value of \(m\) at fixed intervals of time. The results for some specific values of the system's parameters can be seen in Fig. \ref{fig:phases_ex}.

\subsection{Phase diagram}
\begin{figure}[!h]
  \centering
  \includegraphics[width=.45\linewidth]{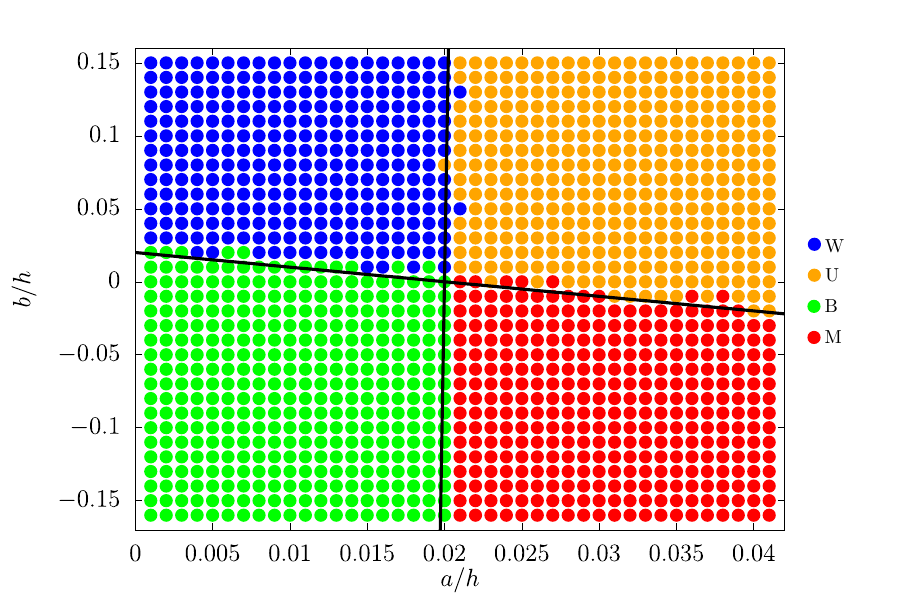}
  \hfill
  \includegraphics[width=.45\linewidth]{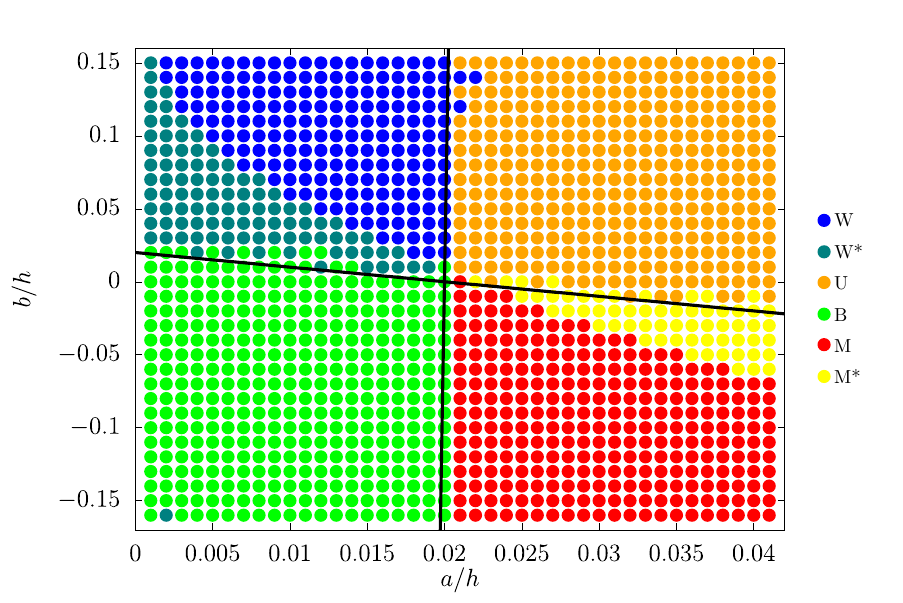}
  \caption{\textbf{Left}: Phase diagram of a system with \(N=50\) generated by a phase detection algorithm and numerical simulations. Each color corresponds to a different phase and black lines correspond to the analytical prediction of the phase boundaries. \textbf{Right:} Phase diagram of a system with \(N=50\) where we introduce a different label \(W^*\) to account for the case in which the central maximum takes a lower value than the maxima at the extremes. Analogously, the label \(M^*\) indicates that the central minimum  takes a higher value than the minima at the extremes.}
  \label{fig:phase_diagram}
\end{figure}

To check the validity of Eqs. \eqref{eq:tr1} and \eqref{eq:tr2} and the phase diagram of Fig. \ref{fig:1} we have carry out simulations for a whole range of values of the parameters of the system (\(a/h\) and \(b/h\)) and then, we have build a phase diagram through a phase-detection algorithm which study the shape of \(P(m)\). In particular, the phase-detection algorithm is based on two simple criteria. First, we study the height of \(P(m)\) at extremes $m=-1, 1$ and compare it with the nearest point $-1+1/N$ and $1-1/N$ to determine whether the function \(P(m)\) is increasing or decreasing. Then, we perform a polynomial fit of the second degree to the vicinity of the middle point $m=0$ and check the sign of the second-degree coefficient to determine if it is a maximum or a minimum. The result for our numerical phase diagram and its comparison with the theoretical one can be checked at Fig. \ref{fig:phase_diagram}. Minor discrepancies between theory and simulations can be imputed to the statistical fluctuations typical of finite simulations of finite systems and the limitations of the phase detection algorithm. 

In the right panel of Fig. \ref{fig:phase_diagram} we include a distinction between two types of behaviors in the W and M phases. We call $W^*$ to the point of the phase diagram which belong to the $W$ phase but has a central maximum of $P(m)$ whose height is lower than the height of the extremes. Analogously, we call $M^*$ to those points of the $M$ phase in which the central minimum of $P(m)$ is higher than the minima at the extremes. This distinction is motivated by a recently published article \cite{llabres2023partisan}, where the partisan noisy voter model is studied. There, the authors find two phase transitions, from $W^*$ to $W$ and from $W$ to $U$ by changing the noise to herding ratio $a/h$, just as we do. 

\begin{figure}[!h]
    \centering
    \includegraphics[width=\linewidth]{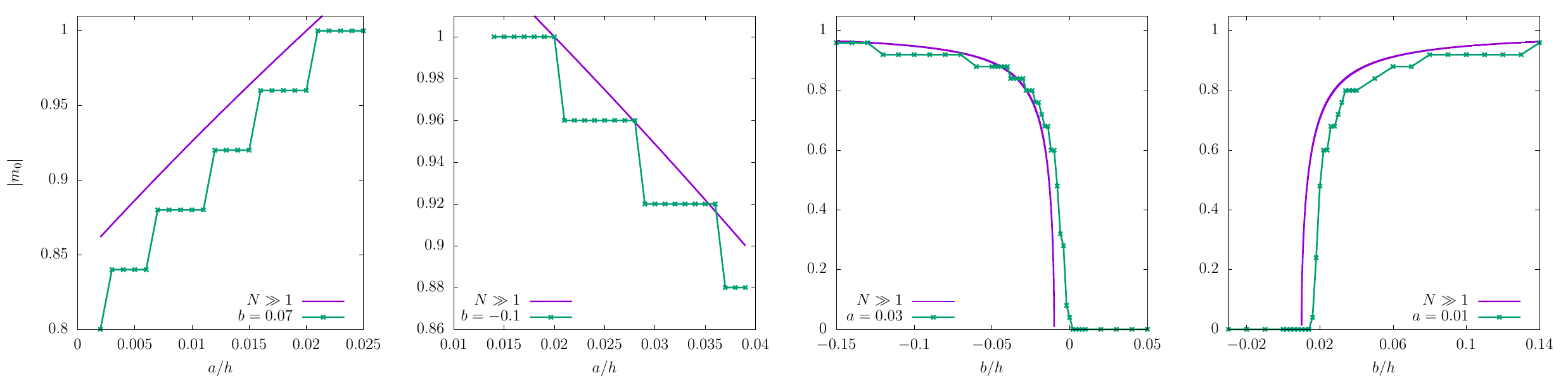}
    \caption{Value of the position of the minimum (first and fourth panels) and maximum (second and third panels) \(m_0\) as a function of the system parameters (\(a/h\) or \(b/h\)) for the different phase transitions. The green points joined by lines correspond to simulation of a system with \(N=50\) agents, and purple continuous lines correspond to the analytical prediction systems with $N\gg1$, Eq.~\eqref{eq:orderp}. From left to right, each panel correspond to the transition: WU, BM, MU and BW.}
    \label{fig:m0}
\end{figure}

\subsection{\(|m_0|\) and moments}
One can use the value of $m_0$ as an order parameter for the transition between the different phases. In Fig.~\ref{fig:m0} we plot the value of $m_0$ obtained from simulations as well as the theoretical prediction from Eq.~\eqref{eq:orderp} for infinite-size, $N\to \infty$, systems . In finite systems, $m$ can only take $2N+1$, values in the range $m_i\in\{-1, -1+1/N, \dots, 0, \dots, 1-1/N, 1\}$. As predicted from the theory for finite systems (see Sect. \ref{sect:max_min}), the function $m_0$ changes from one site $m_i$ to the adjacent site $m_j$ under smooth changes in the system parameters. The transition points are located at the values of $a/h$ and $a/h$ at which $m_0$ takes the value $m_0=1$ (for the two left panels of Fig.~\ref{fig:m0}) or $m_0=1$ (for the two right panels of Fig.~\ref{fig:m0}). 

\begin{figure}[!h]
    \centering
    \includegraphics[width=0.7\linewidth]{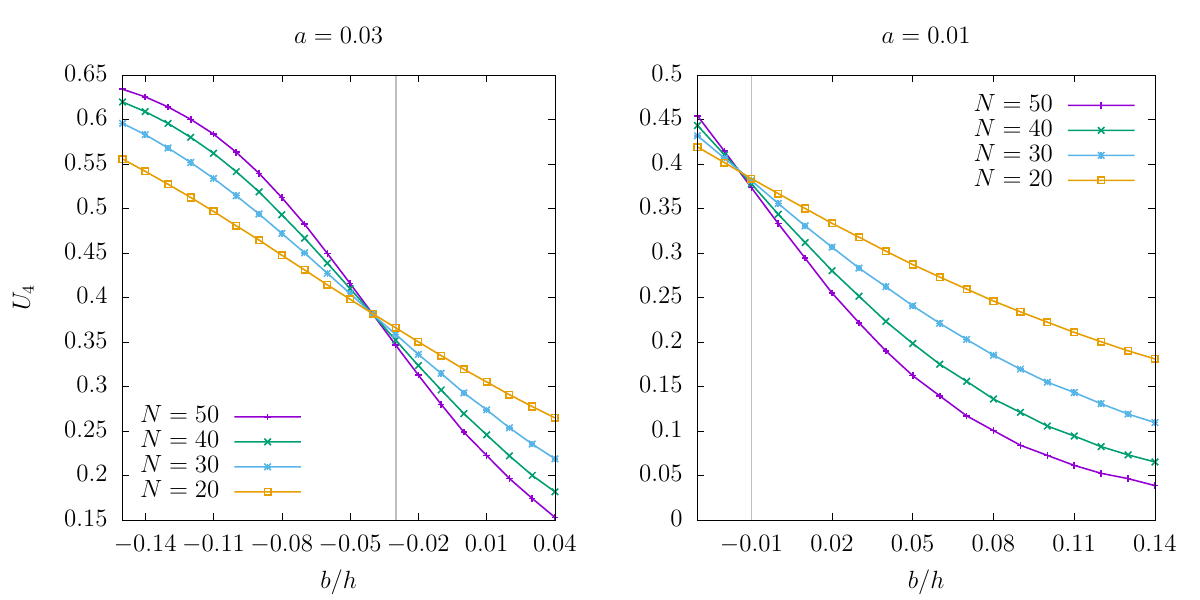}
    \caption{Value of  the Binder cumulant $U_4(N)$ as a function of the system parameter \(b/h\) for different system sizes $N$ for two values of $a/h$. The crossing point in the two  panels signals the presence of a continuous phase transition. The vertical lines are the predictions for $b_c$ from Eq.~\eqref{eq:cl_inf}. }
    \label{fig:binder}
\end{figure}

We can use the moments of $P(m)$ to gain some information about the infinite-size, $N\to \infty$, phase transition, using numerical results from simulations of finite systems. The Binder cumulant \cite{Binder1981_cumulant}, is defined as
\begin{equation}
    U_4(N)=1-\frac{\langle m^4\rangle}{3 \langle m^2\rangle^2}\;, 
\end{equation}
and, following the scaling hypothesis for continuous phase transitions (see, for example,\cite{Amit-Martin,toral2014stochastic,goldenfeld2018lectures}) one would expect it to scale near the critical point (for fixed $h$ and $a$ ) as some function $f_U$ of $b-b_c$,
so that at the critical point $b_c$
\begin{equation}
    U_4(N,b_c)=f_U(0)=\text{cte.}
\end{equation}
Then, at the critical point, $U_4$ takes an $N$-independent value. In Fig.~\ref{fig:binder} we show the value of $U_4$ as computed from numerical simulations. We find that the value of the transition point, obtained from the intersection of curves for different values of $N$ is at good agreement with the analytical prediction from Eq.~\eqref{eq:cl_inf}.

\section{Discussion\label{discussion}}

A new model of opinion dynamics has been proposed and theoretically and numerically studied at the mean-field level. As defined by its rates in Eqs.~\eqref{eq:pimasn} and \eqref{eq:pimenosn}, the model can be regarded as a modification of the noisy voter model (NVM): in addition to the herding term (proportional to the coefficient \(h>0\)) and the intrinsic or noisy term (proportional to \(a>0\)), the model also accounts for the social polarization by including a term proportional to the number of agents holding opposite opinions (proportional to \(b\)). For \(b>0\) agents try to change their opinions with larger polarization, while for \(b<0\) they tend to be more conservative. For \(b=0\), the NVM is recovered. 

By analytically solving the master equation, four different phases have been identified for a finite number of agents, as the parameters of the system change, see Figs.~\ref{fig:1} and \ref{fig:2}. Each of them are characterized by a different form of the steady-state probability function of the fraction of agents holding one of the opinions. Apart from the unimodal and bimodal phases, already present in the NVM, the system can stay in the M and W phases, the latter being a special case of a trimodal phase not observed in \cite{peralta2018analytical}. In the bimodal phase, the herding mechanism dominates over the others, which makes the system spend most of the time close to the consensus states. In the unimodal phase, the noisy mechanism is dominant, and the system is mostly at coexistence.

In the M and W phases none of the mechanisms are globally dominant. In the W phase, close to consensus agents try to copy each others since noise and polarization are smaller than herding. However, when the system is close to coexistence, the polarization (with \(b>0\)) and copying mechanisms compete with each other. This makes the probability distribution to have the W form. Analogously, in the M phase the noise term is dominant except for the regions close to coexistence where agent react against polarization \(b<0\) decreasing the effect of noise and making the coping mechanism more efficient. This gives rise to the M form of the probability function.

All transitions between phases are found to be continuous as approximately given by Eq.~\eqref{eq:orderp}, except the ones through the point \(a/h=1/N,\,b=0\) for which the probability density becomes flat. Moreover, as the number of agents tends to infinity, the bimodal and W phases disappear, meaning that all transitions are of finite character except that between the unimodal and M phases.  

Similar phases (including M and/or W) have been recently found in the noisy voter model with switching environments \cite{caligiuri2023noisy} and the Partisan voter model \cite{llabres2023partisan}. However, in both cases the state of the system is not completely specified by the fraction of agent with a given state, but by something else (the state of the environment and the agent preference, respectively). This suggests that a reduction of the dimensionality of the previous models could produce an effective model with more complex dynamics, similar to the one proposed here. An interesting related question is whether a set of agents can be found in any conceivable state (shape of the probability function) and if the present model can be further generalized to account for all these eventual new states.

\bibliographystyle{elsarticle-num} 
\bibliography{snvm_bib}

\end{document}